# Highly-robust reentrant superconductivity in $CsV_3Sb_5$ under pressure

Xu Chen,[1,+] Xinhui Zhan,[2,+] Xiaojun Wang,[2] Jun Deng,[1] Xiao-bing Liu,[2,*] Xin Chen,[2] Jian-gang Guo,[1,3,*] and Xiaolong Chen[1,3,*]

[1]*Beijing National Laboratory for Condensed Matter Physics, Institute of Physics, Chinese Academy of Sciences, Beijing 100190, P. R. China*

[2]*Laboratory of High Pressure Physics and Material Science (HPPMS), School of Physics and Physical Engineering, Qufu Normal University, Qufu 273100, P. R. China*

[3]*Songshan Lake Materials Laboratory, Dongguan, Guangdong 523808, P. R. China*

[+]These authors contributed equally
[*]xiaobing.phy@qfnu.edu.cn, jgguo@iphy.ac.cn, chenx29@iphy.ac.cn

## Abstract

Here we present the superconducting property and structural stability of kagome $CsV_3Sb_5$ under *in-situ* high pressures. For the initial SC-I phase, its $T_c$ is quickly enhanced from 3.5 K to 7.6 K and then totally suppressed at $P$~10 GPa. Further increasing the applied pressures, an SC-II phase emerges at $P$~15 GPa and persists up to 100 GPa. The $T_c$ rapidly increases to the maximal value of 5.2 K at $P$=53.6 GPa and rather slowly decreases to 4.7 K at $P$=100 GPa. A two-dome-like variation of $T_c$ in $CsV_3Sb_5$ is concluded here. The Raman measurements demonstrate that weakening of $E_{2g}$ model and strengthening of $A_{1g}$ model occur without phase transition as entering the SC-II phase, which is supported by the results of phonon spectra calculations. Electronic structure calculations reveal that exertion of pressure may bridge the gap of topological surface nontrivial states near $E_F$, i. e. $Z2$ invariant. Meanwhile, it enlarges Fermi surface significantly, consistent with the increased carrier density. The findings here point out the change of electronic structure and strengthened electron-phonon coupling should be responsible for the pressure-induced reentrant SC.

## Introduction

A kagome lattice, composed of atoms at the vertices of a two-dimensional network with the corner-sharing triangles, provides a fascinating playground for exploring and studying the novel frustrated, correlated and topological electronic states of matters recently[1,2,3,4]. Due to the special geometry, the kagome lattice naturally possesses Dirac dispersion and nearly flat bands that promote topological and correlation effect [5]. Kagome systems are predicted to host to many exotic electronic states, such as spinless fermions [6,7,8], Mott phase transition [6,9,10], charge density waves (CDW) [6,11,12], and superconductivity (SC) [6,7,13,14]. Up to now, exploring exotic properties, in particular, the interplay between multiple electronic orders in kagome lattices have been challenging.

Recently, a new family of quasi-2D kagome metal $AV_3Sb_5$ (A=K, Rb and Cs) has attracted considerable attentions [15]. The vanadium sublattice of *P6/mmm* $CsV_3Sb_5$ is a structurally perfect kagome lattice. There are two distinct Sb sublattices. The sublattice formed by the Sb1 atom is a simple hexagonal net, centered on each kagome hexagon. All interatomic distances within the kagome layer are equal, as required by the high symmetry of the V1 (Wyckoff 3g) and Sb1 (Wyckoff 1b) sites. The Sb2 (Wyckoff 4h) sublattice creates graphite-like layers of Sb (antimonene) that encapsulate the kagome sheets. The Cs sublattice naturally fills the space between the graphite-like sheets. The superconducting transition temperatures ($T_c$) are 0.93, 0.92, and 2.5 K for three compounds, respectively [16,17,18]. In addition, in normal state they all manifest a proposed charge density wave transitions at 78, 103, and 94 K, respectively [15,19,20]. Interestingly, high-resolution scanning tunneling microscopy (STM) study reveals that such charge order displays a chiral anisotropy [21,22,23], which leads to a giant anomalous Hall effect in absence of resolvable magnetic order or local moments [17,24,25,26], pointing to a precursor of unconventional SC [21,27]. Moreover, $Z_2$ nontrivial topological band structures, including multiple Dirac nodal points near the Fermi level ($E_F$) and possible topological surface states, were demonstrated by angle-resolved photoemission spectroscopy (ARPES) and density-functional theory (DFT) calculations [15,16,18]. Therefore, this V-based kagome metal $AV_3Sb_5$ exhibits multiple coexistences or competitions ordering among SC, electronic correlations and nontrivial band topology.

It is known that *in-situ* pressure tuning is a 'clean' way to tune basic electronic and structural properties without changing the chemical composition, and eventually benefits for elucidating mechanisms of the puzzling state in new superconducting materials [28,29,30,31,32,33]. Very recently, Zhao *et al*. and Chen *et al*. reported that the $T_c$ of $CsV_3Sb_5$ first increases to 7.5 K and quickly reduces under pressure, proving the competition between SC and possible CDW [27,34]. Then, Zhao *et al*. updated the data of relatively high pressure, and the reemergence of SC after 15 GPa was discovered [27,35].

In this study, we performed the *in-situ* high pressure transport measurements on the $CsV_3Sb_5$ sample as pressure is up to 100 GPa. Two superconducting domes in the temperature-pressure phase diagram are revealed. Interestingly, in pressure-induced superconducting state, the $T_c$ and crystal structure are rather robust evidenced by the transport measurements, Raman spectra and phonon spectra calculations. Thus, varied electronic structure and possibly enhanced electron-phonon coupling play important roles in inducing reentrant SC.

**Experimental**

Sample preparation: Single crystals of $CsV_3Sb_5$ were grown from Cs (Alfa, 99.98%), V powder (Alfa, 99.9%) and Sb grains (Alfa, 99.999%) using the self-flux growth method similar to the previous reports [15]. Cs, V and Sb were weighted as the stoichiometric ratio in an argon-filled glove box (A small excess of Cs was used to compensate for volatility of the alkali metal). Then they were packed into $Al_2O_3$ crucible in order and sealed into an evacuated silica tube under partial argon atmosphere. The sealed quartz ampoule was heated to 1273 K and soaked there for 24 h. And then it was cooled down to 823 K at 2 K/h. Finally, the ampoule was taken out from the box furnace and decanted with a centrifuge to separate $CsV_3Sb_5$ single crystals from the flux. The shinning $CsV_3Sb_5$ single crystals were obtained, see Fig. S1.

*In-situ* high-pressure measurements: High pressure resistivity of $CsV_3Sb_5$ single crystals was measured in a physical property measurement system (PPMS, Quantum Design) by using a diamond anvil cell (DAC) at temperatures of 2–300 K. We used the van der Pauw method for electrical transport measurements on the $CsV_3Sb_5$ samples. Pe–Cu cells were used for the resistance experiments. The cubic boron nitride (cBN) powders (200 and 300 nm in diameter) were employed as pressure medium and insulating material. The pressure was measured using the ruby fluorescence method at room temperature each time before and after the measurement. In addition, in-situ Raman spectra under high pressure were taken (Horiba, Lab- RAM HR revolution) to characterize their structures.

Theoretical calculations: We performed structural optimization and electronic property calculations within framework of the DFT+U by the Vienna ab initio Simulation Package (VASP) code [36,37,38,39]. The generalized-gradient approximation (GGA) is used for the exchange-correlation term within the Perdew-Burke-Ernzerhof (PBE) approach [40,41,42]. The additional Hubbard-like term is used to treat the on-site Coulomb interactions on the localized *d*-states of vanadium atoms. A plane wave energy cutoff of 350 eV and the Monkhorst–Pack *k*-meshes with a maximum spacing of $2\pi \times 0.15$ Å$^{-1}$ were used to ensure that all the enthalpy calculations are well converged to better than 1 meV/atom. To examine the dynamical stability of the structure, we performed phonon calculations by using the finite displacement method as implemented in the PHONOPY code [43]. The Brillouin zone was sampled with a $3 \times 3 \times 2$ supercell. Crystal orbital Hamilton population (COHP) analysis was performed using the LOBSTER package [44,45] in conjunction with VASP.

## Results and discussion

In the inset of Fig. 1(a), one can see that the temperature dependent resistivity of $CsV_3Sb_5$ under ambient pressure, in which the kink due to proposed charge density wave of 90 K and SC of 3.5 K coexist. The residual resistivity ratio value is about 25, indicating good quality of our sample. In a typical *in-situ* high pressure measurement (Run #1), two clear features emerge as shown in the main panel of Fig. 1a. Firstly, the $T_c^{onset}$ of SC-I phase increases to 7.6 K at 0.7 GPa and then decreases to below 2 K at 9.0 GPa, leading to a dome-like $T_c$. This pressure-dependent $T_c$ is consistent with the recent reports [27,34]. The *H*-dependent resistivity at 1.5 GPa are plotted in Fig. 1b. It is found that the *H* of 3.0 T can totally suppress the superconducting transition. The upper critical field $\mu_0H_{c2}$, 4.2 T, is obtained based on the linear fitting. Another feature is that the magnitude of resistivity and the slops of $\rho$-*T* curve continuously decrease. The former one is due to the pressure-induced compact of crystal, and the latter may relate to the variation of electron-phonon coupling under pressure. In Fig. 1c, we can find that the sample is normal metal without any signature of SC in the pressure range of 12.0–14.9 GPa. The whole magnitude of resistivity and residual resistivity increases with *P*. As *P*>19.5 GPa, a new superconducting phase appears. The $T_c^{onset}$ of reentrant SC is 2.8 K, which slowly increases to $T_c$ of ~5.1 K at ~57.1 GPa. One thing should be noted is that the SC phase survives after the pressure is released, indicating a reversible structure even the pressure is squeezed up to 57.1 GPa, see Fig. S2. In the Run #2, we further increase the pressure to 100 GPa. In Fig. 1d, it is found that the second SC re-enters as *P*>20 GPa, exhibits peak $T_c^{onset}$ of 5.2 K at 53.6 GPa and then gradually decreases. The $T_c$ is rather robust against pressure. At 100 GPa, the $T_c^{onset}$ is 4.7 K. We measured the *H*-dependent superconducting transition under *P*=53.6 GPa, and plotted the curves in Fig. 1e. The SC is suppressed as *H*>2 T. The calculated $\mu_0H_{c2}(0)$ is about 3.5 T. We show the temperature-dependent $\mu_0H_{c2}$ of $CsV_3Sb_5$ under selected pressure. The pressure dependence of $\mu_0H_{c2}$ are plotted in Fig. 1f and Fig. S3, and the values are smaller than the Pauli limit of $1.84T_c$.

The pressure-dependent $T_c$ and related physical parameters are summarized in Fig. 2. There are two dome-like superconducting states under pressure as shown in Fig. 2a. In the first SC range (SC-I), applying pressure rapidly enhances the $T_c$ to ~7.6 K and then suppresses the $T_c$ below 2 K as *P*~10 GPa. In the second SC range (SC-II), as *P*>15 GPa, the $T_c$ continuously increases to the highest $T_c$ of 5.2 K, which is a little lower than that in SC-I region. The two SC are separated by ~10 GPa. As previously reported, the highest $T_c$ of reentrant SC is higher than that in SC-I phase such as FeSe-based superconductors. In $K_xFe_2Se_2$ and $(Li_{1-x}Fe_x)OHFe_{1-y}Se$ superconductors, the SC-I phase of 30 K and 41 K are rapidly suppressed at *P*=10 GPa and 4 GPa, respectively. Then pressure-induced SC-II phase emerge with the maximum $T_c$ of 48.7 K and 50 K [32,46,47]. According to the transport property analysis, the suppression of first SC phase is related to a possible quantum phase transition from Fermi liquid to a non-Fermi liquid behavior [48]. In some superconductors like $KMo_3As_3$, the $T_c$ of pressure-induced SC-II phase is a little lower than that of initial SC-I phase [49].

In the normal state, the pressure-dependent key transport parameters are plotted in Fig. 2b-f. We fit the $\rho(T)$ curve by the equation $\rho=\rho_0+AT^\alpha$ from $T_c$ to 50 K. Here $\rho_0$ is residual resistivity, and A constant. The pressure-dependent index $\alpha$ is plotted in Fig. 2b. In the SC-I, the $\alpha$ increases from 1.69 to 1.93, then it continuously decreases to 1.50 as the reentrant SC occurs. As the $P$ goes through the optimal value, the $\alpha$ increases again to 1.87. Such behaviors imply that the electron-correlated states possibly transit from Fermi-liquid to non-Fermi-liquid state, and then back to Fermi-liquid state again [50,51]. For the A, in SC-I region, it quickly increases to the peak value at 0.7 GPa, and then lowers to the minimal as the $T_c$ is totally suppressed, see Fig. 2c. In the NSC and SC-II regions, the A shows peak value of 0.029 $\mu\Omega$ cm/K$^{-\alpha}$. On the other hand, in a pressurized sample, the $\rho_0$ generally decreases as increasing $P$ as the trend in SC-I region. But the enhancement of $\rho_0$ in higher $P$ is unusual, which may be caused by disorder if there is no metal-insulator transition.

Based on the Bloch-Grüneisen model of scattering of electron by longitudinal acoustic vibrations [52]:

$$\rho(T) = \rho_0 + B \times \frac{T^n}{\Theta_D^{n-1}} \int_0^{\Theta_D/T} \frac{z^n dz}{(e^z-1)(1-e^{-z})} - kT^3$$

Where $\rho_0$ is residual resistivity, the value of $n$ is typically fixed to be 3, $k$ the cofficient of the cubic term, $\Theta_D$ the Debye temperature from the resistivity data and $B$ a prefactor depending on the material. We fitted the $\rho(T)$ data above $T_c$ at each $P$, and obtained the $\Theta_D$ as shown in Fig. 2d. It can be found that the $\Theta_D$ firstly rapidly increases to the maximal 344 K in SC-I range, and then slowly decrease to a constant value of 150 K as entering the SC-II range. In addition, we measure the pressure-dependent carrier density $n$ as shown in Fig. 2e, and find that it gradually increases from $1.6\times10^{22}$ cm$^{-3}$ to $2.9\times10^{22}$ cm$^{-3}$. The magnitude of $n$ is consistent with the reported values in analogous compounds [24,25].

It is instructive to check the structural stability at high pressure, thus we calculated phonon dispersions and projected phonon density of states at $P$=10 GPa, 20 GPa, 40 GPa, and 60 GPa, see Fig. S4. The optimized crystallographic parameters are summarized in Table S1. No imaginary frequencies are observed, indicating the dynamical stability of the *P6/mmm* structure under pressure. Moreover, the contributions of phonon dispersions mainly come from the vibration of V and Sb atoms. The heavier Sb atoms dominate the low-frequency branches, while the relatively light V atoms contribute significantly to the high-frequency modes. Cs vibration has a smaller contribution over the whole frequency range due to the negligible chemical bonding with Sb atoms. More interestingly, it is found from Fig. S4 that the low-energy acoustic phonons are softening around M point with increasing pressure up to 40 GPa. On the other hand, according to the structural symmetry, three Raman-active models assigned as $A_{1g}$, $E_{2g}$ and $E_{1g}$ due to vibration of Sb are drawn in Fig. 3a. The positions of three peaks can be theoretically estimated from the calculated phonon spectra at 10 GPa, 20 GPa, 40 GPa and 60 GPa. One can find that three peaks are blue-shifted under applied pressure as shown in Fig. S5. We measured the pressure-dependent Raman spectra of $CsV_3Sb_5$ from 1 to 53 GPa and plotted them in Fig. 3b, Fig. S6 and S7. At low $P$, two peaks of 119.5 cm$^{-1}$ and 135.3 cm$^{-1}$ are observed, which belong to $E_{2g}$ and $A_{1g}$ models associated with longitudinal and transversal vibration of Sb atoms. As $P>40$ GPa, a small peak at ~100 cm$^{-1}$ emerges, which is $E_{1g}$ model due to in-plane relative vibration of upper Sb and lower Sb atoms. As increasing $P$, the peaks of $E_{2g}$ and $A_{1g}$ indeed move to higher wave number. Simultaneously, the intensity of $E_{2g}$ slowly decreases while that of $A_{1g}$ increases. This anomalous change leads to that the intensity ratio of $A_{1g}/E_{2g}$ starts to increase as $P>20$ GPa and then reaches the peak value at $P=30$ GPa, see Fig. 3c. From above lattice dynamic information, we can deduce that the initial structure of $CsV_3Sb_5$ with space group of $P6/mmm$ is very robust, and no structural transition is detected as $P=53$ GPa. Intuitively, the reentrant SC may relate to the weakened vibration of $E_{2g}$ model, which may induce partial strengthened electron-phonon coupling.

We calculated the electronic band structures at various pressures, as shown in Fig. 4. It can be seen the electronic states of $CsV_3Sb_5$ near the $E_F$ are mainly attributed by the V-$d$ and the Sb-$p$ orbitals. With the increase of pressure, the density of states (DOS) near $E_F$ decreases from 5 to 3 states/eV per formula, see the right panel of Fig. 4a-d and Fig. S8. Meanwhile, applying of pressure lifts the energy of middle band (blue color), closing the continuous direct gap in partial Brillouin zone. It implies that the $Z_2$ topological invariant may disappear at $P>20$ GPa [53,54], see Fig. 4c and d. In addition, the broadening of the band structure leads to more bands crossing through the $E_F$ especially along Γ-A-L direction. From the pressure-dependent topology of Fermi surface (Fig. 4e-h), it can be found that the scale of Fermi surface increases, consistent with the enhanced carrier density shown in Fig. 2e. Strikingly, in Fig. 4h, the $P6/mmm$ structure exhibits Fermi surface nesting at 40 GPa along Γ-A and M-L directions, associated with highly dispersive bands in this direction as discussed above. Therefore, we suggest that the nested Fermi surface arising under pressure induces phonon softening that strengthens the electron–phonon coupling, favoring the reemergence of high-$T_c$ SC in pressurized $CsV_3Sb_5$.

From Fig. S9a, it is found that the lattice parameter $c$ reduces rapidly in the low-pressure region and gradually changes slowly with increasing the pressure over 40 GPa. In contrast, the lattice parameter $a$ does not change significantly, leading to the slow decrease of interatomic distances within the kagome layer. Especially after 60 GPa, the bond length of V-Sb1 (V-V) exhibits a tendency to exceed that of V-Sb2 under pressure, see Fig. S9b. In order to further investigate the bonding character of $CsV_3Sb_5$, we calculated projected crystal orbital Hamiltonian population (pCOHP) and plotted them in Fig. S10. It can evaluate the weighted population of wave functions on two atomic orbitals of a pair of selected atoms. It is noted that the occupied bonding states in $CsV_3Sb_5$ at 40 GPa move to deeper energy below the $E_F$, indicating higher stability of $P6/mmm$ $CsV_3Sb_5$ at 40 GPa. Moreover, in Fig. S11, we use the integrated COHP (ICOHP) to quantitatively estimate the bonding strength. It is noted that the V-Sb1 covalent bonding in kagome net of vanadium are the strongest, and with increasing pressure, the V-Sb1 bond strengthens not significantly as that of V-Sb2 and V-V bonds, suggesting the stable kagome layer upon compression as verified above.

## Conclusions

In this work, we have investigated the SC, transport property and structural stability of CsV$_3$Sb$_5$ under high pressure up to 100 GPa. Initial SC phase-I is rapidly suppressed at $P\sim$10 GPa, and then a second SC phase-II emerges as $P$>15 GPa. Interestingly, the SC-II exhibits a dome-like $T_c$, in which the $T_c$ quickly increases to peak $T_c^{onset}$ of 5.2 K and then slowly decreases to 4.7 K at $P$=100 GPa. Theoretical calculations and Raman measurements demonstrate that the initial crystal structure can persist in the whole pressure range. Therefore, the reentrant SC should relate to variation of electronic structure and enhanced electron-phonon coupling due to partial phonon softening. The findings here suggest the highly stable SC and structure in such kagome lattice deserves further investigation.

## Acknowledgements

This work is financially supported by the National Key Research and Development Program of China (No. 2017YFA0304700, 2018YFE0202601 and 2016YFA0300600), the National Natural Science Foundation of China (No. 11804184, 11974208, 51532010 and 51772322), the Chinese Academy of Sciences under Grant QYZDJ-SSW-SLH013, the Shandong Provincial Natural Science Foundation (ZR2020YQ05, ZR2019MA054, 2019KJJ020).

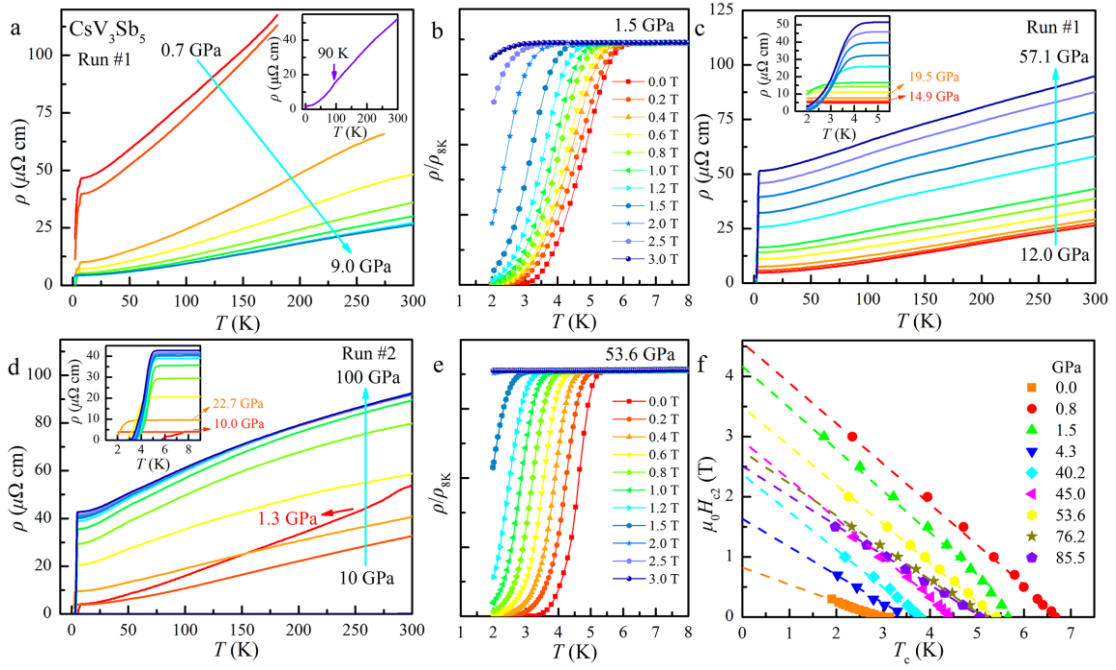

**Figure 1.** Transport Properties of CsV$_3$Sb$_5$ at various pressures. (a) $\rho$-T curves at temperature range of 2-300 K for 0.7-9.0 GPa. The inset of (a) is $\rho$-T curve for 0 GPa. (b) $\rho/\rho_{8K}$ around $T_c$ under external magnetic field at 1.5 GPa. (c) $\rho$-T curves at temperature range of 2-300 K for 12.0-57.1 GPa. The inset of (c) is $\rho$-T curves at temperature range of 2-6 K. (d) $\rho$-T curves at temperature range of 2-300 K for 1.3-100 GPa. The inset of (d) is $\rho$-T curve at temperature range of 2-6 K. (e) $\rho/\rho_{8K}$ around $T_c$ under external magnetic field at 53.6 GPa. (f) Temperature dependence of the upper critical field $\mu_0H_{c2}$ at different pressures. The broken lines represent the linear fitting curves.

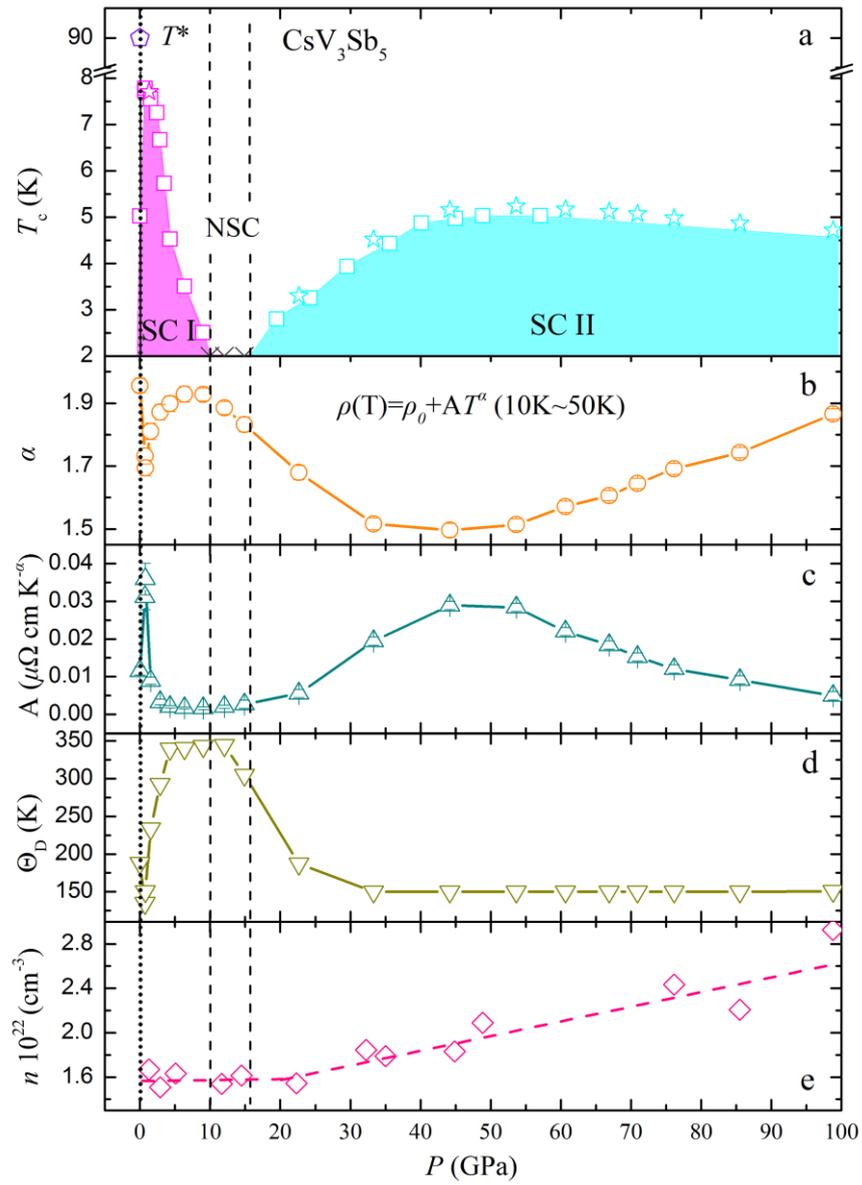

**Figure 2.** Pressure dependence of (a) $T_c$ and related physical parameters (b) index $\alpha$, (c) prefactor A, (d) Debye temperature $\Theta_D$ and (e) carrier density $n$ of $CsV_3Sb_5$.

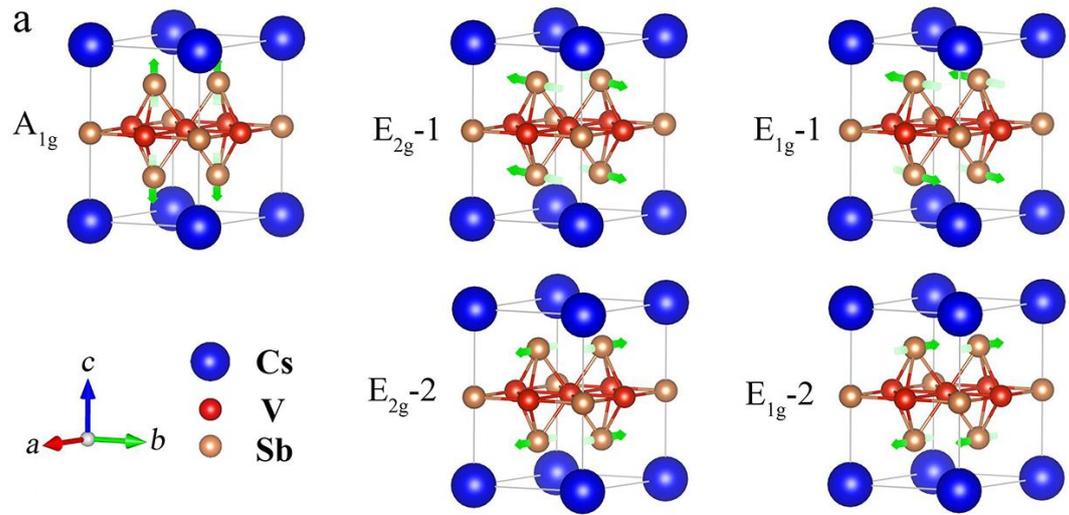

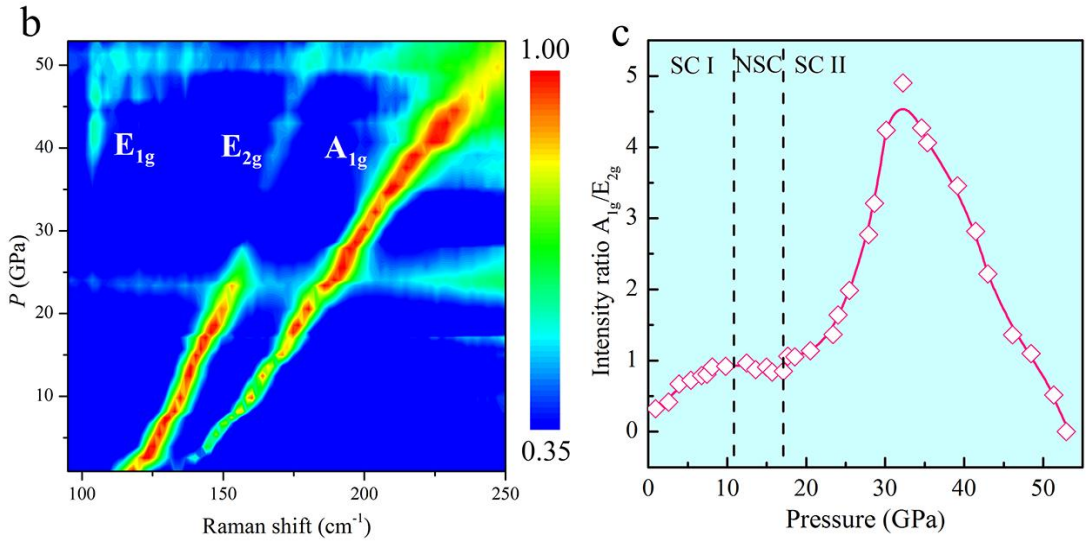

**Figure 3.** (a) Three Raman-active models of $CsV_3Sb_5$ as $A_{1g}$, $E_{2g}$ and $E_{1g}$. (b) Pressure-induced Raman intensity and position change of $CsV_3Sb_5$ in the pressure range of 1-53 GPa. (c) Intensity ratio of $A_{1g}/E_{2g}$ taken from (b).

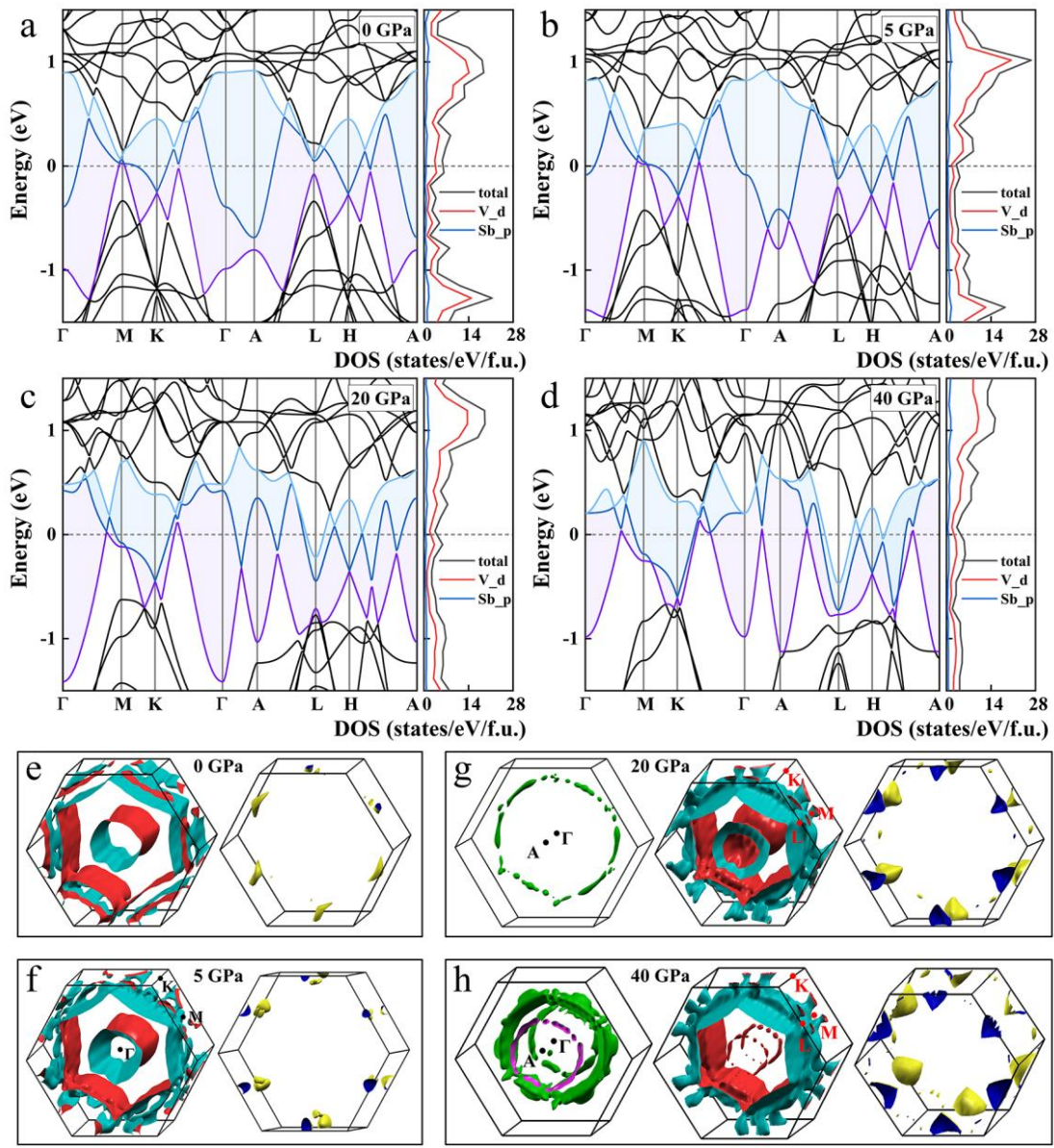

**Figure 4.** Electronic band structures and partial density of states (PDOS) for CsV$_3$Sb$_5$ at (a) 0 GPa, (b) 5 GPa, (c) 20 GPa and (d) 40 GPa, respectively. Three bands crossing the $E_F$ are plotted with different colors (light blue, blue and purple). Fermi surface of CsV$_3$Sb$_5$ at (e) 0 GPa, (f) 5 GPa, (g) 20 GPa and (h) 40 GPa, respectively.